# Nebular Analysis of PNe NGC 3242

## Soubhik Chatterjee[1]

[1]Institute of Radio Physics and Electronics, University of Calcutta

**ABSTRACT**

*The work presented forms a part of the IASc-INSA-NASI Summer Research Fellowship (SRFP) project on the physical & chemical properties of planetary nebulae. Spectral observations of NGC3242 recorded using the VBT Observatory (IIA), Kavalur (TN) for orthogonal slit positions are used for the study. The spectral data were reduced and analyzed using the IRAF data analysis package following the standard procedure to obtain the one dimensional spectrum. Nebular Analysis is performed on the same using NEAT (Wesson et al, 2012) to obtain elemental abundances, densities and temperatures for the PNe and the results are compared to existing literature.*

## 1 INTRODUCTION

Planetary Nebulae are isolated nebulae. Contrary to their names, they have nothing to do with planets, and instead are the remnants of a dying intermediate mass star's last gasp. When a sun-like star of intermediate mass exhausts its nuclear fuel, it undergoes a series of dramatic transformations, eventually shedding its outer layers in a violent outpouring of gas and dust. These expelled layers form a luminous shell, often intricately shaped and adorned with vivid colors, as a result of ionized gasses emitting light due photoionisation from the central star.

PNe often poses a bilateral symmetry. While the central star contracts and rapidly evolves into a dense & hot white dwarf, the surrounding planetary nebula expands into space with velocities several times the speed of sound. As they are decreasing in density over time, there is a decrease in emission. Thus on a cosmic timescale, PNe become rapidly unobservable, having a mean lifetime of 10,000 years. Typical densities of PNe range from 100 to 10000 particles of matter per cubic centimeter, with an overall mass ranging from 0.1 $M_\odot$ to 1$M_\odot$. Space distribution and kinematic properties of PNe indicate them being fairly old (Population-I) objects indicating that the bulk of PNe observed, though new, are near terminal stages of stellar evolution.

## 2 OBSERVATION & DATA REDUCTION

Low-Medium resolution optical spectroscopic observations of PNe NGC3242 were made using the Opto-mechanics research (OMR) grating spectrograph attached with the 2.34m VBT located at Vainu Bappu Observatory (IIA), Kavalur in Tamil Nadu, India. Observations were made with clear sky conditions on February 26, 2021 and March 05, 2022 for orthogonal slit positions, with the telescope guided by an auto guider to overcome any drift in the sky. The slit of the spectrograph has a length of 2.8′ and its width was set at 2″. The slit position was chosen such that it covers the whole optical nebula and passes through the central star, a required condition to accurately constrain the nebular parameters. The observational data was already available at the start of this study, and kindly provided by Prof. Muthumariappan C, Indian Institute of Astrophysics (Bangalore).

The observations were recorded on an Andor DU 920P-BEX2-DD-9998 CCD having a panel of 1024px x 256px, of pixel size 26μm x 26μm. The CCD has read noise of 4.31e with a gain of 2.33e/ADU. A grating of 600 lines/mm has been used and the spectra recorded for two wavelength region settings. The blue region setting has a wavelength coverage of 3800Å to 6300Å, and the red region setting has a coverage of 5800Å to 8200Å. The wavelength dispersion across a pixel is 2.91Å, providing a spectral resolution of 1500 at 4400Å.

The spectral data provided for reduction & analysis was in the form of FITS- Flexible Image Transport System, each consisting of a header section containing the metadata, and the raw spectrum stored in the form of arrays. For purposes of reduction to account for the device (CCD) signature and spectrum extraction, the object frames are subject to the standard reduction procedure using the IRAF Community Version. The one dimensional spectrum was thus obtained. There existed an overlap of spectral regions between the spectrum for the two wavelength settings used. For purposes of scaling the line fluxes, the strength of emission lines in the overlapping spectral region & exposure time of the target frame has been taken into account. The absolute flux for each line is measured by fitting a gaussian to the emission lines and summing up the area under the curve. The spectra lines are normalized to a flux scale such that F(Hβ)=100, with Hβ corresponding to the beta transition line of the Hydrogen Balmer series at 4861Å.

## 3 SPECTROSCOPY OF PLANETARY NEBULAE

PNe are difficult to study owing to their low luminosities compared to bright H II regions of Diffuse Nebulae. The source of energy that enables the nebulae to radiate is almost always the UV radiation from stars within them. One or more hot stars are associated with the nebulae, having an effective T★ greater than 30,000K, and the UV photons emitted by them transfer energy to the nebula. Due to higher T★ of the associated (central) star, typical PNe are generally more highly ionized compared to H II regions. Thus a PNe spectrum is expected to include not only the recombination lines of H I and He I, but also of species such as He II.

Collisions between thermal electrons and ions excite the low lying energy levels of the ions. Downwards radiation transitions from these excited levels have very small transition probabilities. However at the very low densities seen with PNe, collisional de-excitation is even less probable. Thus almost every excitation leads to the emission of a photon. Excitations between such excited levels and the ground levels are forbidden by the electric dipole selection, and thus occur by magnetic dipole and/or electric quadrupole transitions. Such excitations result in "forbidden lines". The spectrum of a PNe is also expected to include such forbidden lines.

On the basis of ionization, the structure of a PNe can be divided into three zones- (i) high ionization zone, consisting of the innermost regions of the PNe closest to the central star having highly ionized species which indicate the presence of very energetic processes due to the intense radiation, (ii) medium ionization zone consisting of a transition region of intermediate radiation intensities, and (iii) a low ionization zone dominated by neutral atoms and low ionization species situated farthest radially from the central star. The low ionization region may extend well into the surrounding interstellar medium where interactions between the ejected materials and stellar winds take place.

In the IR region, most nebulae have a strong continuous spectrum due to the dust particles present being heated, as they compete with the existing atoms, ions and molecules for radiation emitted by the central star of PNe. Thus a stronger IR spectrum indicates the presence of a relatively higher amount of dust.

## 5 RESULTS

In this section, the results obtained from the spectrum and the nebular diagnostics for NGC3242 are presented. A comparative study of the derived results with the existing literature has been made.

The resultant line fluxes obtained from the spectral observations are subjected to NEAT (Wesson et al, 2012), and the observed fluxes along with the de-reddened fluxes and abundances are presented (TABLE-1). A mean interstellar extinction coefficient, c(Hβ) of 0.156 has been calculated for the observations. A comparison between the flux observed in this study for prominent emission lines and those available in the literature is also presented (TABLE-2).

The results associated with the various nebular parameters are presented for elemental abundances (TABLE-3), and density & electron temperatures (TABLE-4). A comparison has been made between the resulting work and existing literature on NGC3242 with varying slit positions and apertures.

While this study is based on spectroscopy with mutually orthogonal slit position, comparisons are made with the studies of Monteiro et al (2013) covering the entire nebula, Tsamis et al (2003) using a single long slit of width 2'', and Krabbe & Copetti (2006) using data from different sources having different aperture/slit sizes.

**TABLE-1.** Observed & De-reddened flux along with abundances for NGC 3242

| Emission Line | Wavelength (Å) | Observed Flux (F) | De-reddened Flux (I) | Abundance |
|---|---|---|---|---|
| H I | 3835 | 2.560 | 2.787 | - |
| [Ne III] | 3868 | 74.80 | 81.244 | 2.977E-05 |
| He I | 3888 | 16.500 | 17.896 | 1.204E-01 |
| [Ne III] | 3967 | 55.800 | 60.176 | 7.319E-05 |
| He I | 4026 | 1.260 | 1.353 | 5.464E-02 |
| [S II] | 4068 | 0.958 | 1.025 | 1.216E-07 |
| Hδ | 4101 | 34.200 | 36.502 | - |
| Hγ | 4340 | 49.500 | 51.802 | - |
| [O III] | 4363 | 14.000 | 14.623 | 1.325E-04 |
| He II | 4541 | 0.859 | 0.883 | 2.231E-02 |
| N III | 4640 | 17.700 | 18.048 | 4.682E-03 |
| He II | 4686 | 24.900 | 25.289 | 2.170E-02 |
| [Ar IV] | 4711 | 6.060 | 6.141 | 7.240E-07 |
| [Ar IV] | 4740 | 6.120 | 6.186 | 7.240E-07 |
| Hβ | 4861 | 100.00 | 100.001 | - |
| [O III] | 4958 | 375 | 371.793 | 1.527E-04 |
| [O III] | 5007 | 916 | 904.350 | 1.245E-04 |
| He II | 5411 | 1.08 | 1.030 | 1.127E-02 |
| He I | 5876 | 14.00 | 12.963 | 7.800E-02 |
| [O I] | 6300 | 00.365 | 0.330 | 2.199E-07 |
| [S III] | 6312 | 0.521 | 0.471 | 3.640E-07 |
| [N II] | 6548 | 125 | 111.528 | 3.370E-05 |
| Hα | 6563 | 323 | 287.973 | - |
| N II | 6610 | 8.27 | 7.355 | 3.088E+00 |
| He I | 6678 | 4.37 | 3.874 | 8.278E-02 |
| He I | 7065 | 3.88 | 3.377 | 6.276E-02 |
| [Ar III] | 7135 | 10 | 8.675 | 4.198E-07 |
| [Ar III] | 7751 | 2.35 | 1.988 | - |
| [Cl IV] | 8045 | 0.519 | 0.434 | 1.876E-08 |
| He II | 8236 | 0.560 | 0.466 | - |

**TABLE-2.** Total line fluxes obtained relative to Hβ and compared to those obtained in literature

| Emission line | Wavelength (Å) | Observed Flux | Monteiro et al. (2013) | Tsamis et al. (2003) |
|---|---|---|---|---|
| [Ne III] | 3868 | 74.80 | - | 98.50 |
| Hδ | 4101 | 34.200 | - | 25.5 |
| Hγ | 4340 | 49.500 | 43.40 | 46.10 |
| [O III] | 4363 | 14.000 | 13.00 | 13.20 |
| N III | 4640 | 17.700 | 4.7 | 1.6 |
| He II | 4686 | 24.900 | 10.90 | 25.50 |
| [Ar IV] | 4711 | 6.060 | 6.7 | 4.9 |
| [Ar IV] | 4740 | 6.120 | 4.7 | 4.5 |
| Hβ | 4861 | 100.00 | 100.00 | 100.00 |
| [O III] | 4958 | 375 | - | 427 |
| [O III] | 5007 | 916 | 1287 | 1300 |
| He I | 5876 | 14.00 | 18 | 12.20 |
| [O I] | 6300 | 00.365 | 0.03 | 0.05 |
| [S III] | 6312 | 0.521 | 0.80 | 0.70 |
| Hα | 6563 | 323 | 437 | 310 |
| He I | 6678 | 4.37 | 3.90 | 3.40 |

**TABLE-3.** Abundances obtained from integrated line fluxes compared to those in literature

| This Work | | Monteiro et al. (2013) | Tsamis et al. (2003) | Krabbe & Copetti (2006) |
|---|---|---|---|---|
| Ion | Abundance | Abundance (ICF) | Abundance (ICF) | Abundance (ICF) |
| He/H | 0.1005 | 0.09 | 0.10 | 0.10 |
| N/H | 3.370E-05 | 2.05E-05 (80.0) | 3.41E-05 (1.53) | 4.19E-05 (220) |
| O/H | 1.327E-04 | 2.55E-04 (1.07) | 3.31E-04 (1.17) | 3.09E-04 (1.30) |
| Ne/H | 4.832E-05 | - | 7.80E-05 (1.18) | 6.46E-05 (1.31) |
| Ar/H | 1.144E-06 | - | 9.79E-07 (1.01) | - |
| S/H | 1.216E-07 | 2.24E-06 (3.00) | 2.40E-06 (3.52) | 2.60E-06 (4.20) |
| Cl/H | 1.876E-08 | - | 8.77E-08 (3.59) | - |

**TABLE-4.** Density & Temperatures analyzed using NEAT compared to those in literature

| Diagnostic (cm$^{-3}$/K) | This work | Monteiro et al. (2013) | Tsamis et al. (2003) | Krabbe & Copetti (2006) |
|---|---|---|---|---|
| $N_e$ [Ar IV] | 3,364.30 | 2,640 ± 400 | 3,040 | 3,665 ± 141 |
| [Ar IV] 4740/4711 | 1.008 | - | 0.92 | - |
| $T_e$ [O III] | 13,523.10 | 12,900 ± 720 | 11,700 | 12,140 ± 31 |
| [O III] (4959+5007)/4363 | 87.880 | - | 4.23 | - |

## 6 CONCLUSION

The data reduction & one dimensional spectral extraction for recorded frames of PNe NGC3242 for mutually orthogonal slit position has been successful. The presence of emission lines such as those of high ionization species of [Ar IV] at 4711Å & 4740Å, and of forbidden lines of species [Ne III] at 3868Å & 3967Å, [N II] at 6548Å, & [O III] at 4363Å, 4958Å & 5007Å among others attests to the high ionization & the presence of forbidden mechanism as has been discussed in the preceding section. On investigation of the physical and chemical properties from the observed spectrum of NGC 3242, it can be concluded that the values arrived at are consistent with those available in the literature.

A further attempt at the 1D dusty photoionization modeling of NGC3242 can be made using Cloudy, with the results of this study, taking into account various other constraints such as the involvement of grains of varying sizes, & distribution, the extent of grain heating, among others.

## 7 ACKNOWLEDGEMENT

## 8 REFERENCES

A. C. Krabbe, M. V. F. Copetti, Chemical abundances in seven galactic planetary nebulae, A&A 450 (2006)
C. Rola, D. Petat, On the estimation of intensity for low S/N ratio narrow emission lines, A&A 287 (1994)
D. E. Osterbrock, Gary J. Ferland, Astrophysics Of Gaseous Nebulae and Active Galactic Nuclei (2006)
George B. Rybicki, Alan P. Lightman, Radiative Processes in Astrophysics (2004)
H. Monteiro, D. R. Gonçalves, M. L. Leal-Ferreira & R. L. M. Corradi, Spatially resolved physical and chemical properties of the planetary nebula NGC 3242, A&A 560, A102 (2013)
R. Wesson, D. J. Stock, P. Scicluna, Understanding and reducing statistical uncertainties in nebular abundance determinations, MNRAS, 422(4) pp.3516–3526
Y. G. Tsamis, M. J. Barlow, X.-W. Liu, I. J. Danziger, P. J. Storey, A deep survey of heavy element lines in planetary nebulae – I. Observations and forbidden-line densities, temperatures and abundances, MNRAS, 345(1) pp.186–220